\documentclass[preprint,showpacs,preprintnumbers]{revtex4}

\usepackage{amssymb}
\usepackage{amsfonts}
\usepackage{graphicx}
\usepackage{dcolumn}
\usepackage{bm}
\usepackage[dvipdfm]{hyperref}

\begin{document}

\title{A nanoscale window for probing Planck scale phenomena}
\author{Shi-Dong Liang}
\altaffiliation{Email: stslsd@mail.sysu.edu.cn}
\affiliation{State Key Laboratory of Optoelectronic Material and Technology, and
Guangdong Province Key Laboratory of Display Material and Technology, School
of Physics, Sun Yat-Sen University, Guangzhou, 510275, People's Republic of
China}
\date{\today }

\begin{abstract}
The noncommutative space provides a framework to understand phenomena in
Planck scale physics. However, there is no any direct experimental
evidence to demonstrate the existence of noncommutative space. We propose an
experimental scheme based on the Aharonov-Bohm effect in the nano-scale
quantum mechanics to probe the phenomena in the noncommutative phase space. 
By the Seiberg-Witten map, the free electrons of the nano-scale ring in noncommutative phase
space can be mapped equivalently to the quantum mechanical (Heisenberg's algebra) phase
space with an extra effective magnetic flux. We introduce two variables related to the
persistent current in the ring to probe the noncommutative phase space
effect. We give a value-independent criterion to detect the existence of the
noncommutative phase space. Namely the answer for existence or nonexistence
of the noncommutative phase space depends only on the trend of the curves,
rather than the values of the observation data. It can be expected to catch
the noncommutative phase-space effect by this scheme.
\end{abstract}

\maketitle



\section{Introduction}
When physicist horizon extended from the macroscopic scale to the Planck
scale create a weird world in the Planck scale, noncommutative geometry and
algebra.
The fundamental difficulties in theoretical physics, such as the singularity
in particle physics, nonlocality and dark energy in quantum gravity and
quantum cosmology, hint the possibility of existence of the finite length
and time (Planck scale), by which physicist can understand consistently all
phenomena in physics world.\cite{Douglas} 
The minimum length and time leads to noncommutative spacetime, namely
noncommutative algebra and geometry rules the Planck world. What is the
noncommutative spacetime ?
The noncommutative concepts in physics can be traced back to the angular
momentum algebra in classical mechanics. These concepts are upgraded to any
pair of conjugate variables called Heisenberg's algebra in quantum mechanics.
It infers that the elementary particles have an intrinsic uncertainty and
nonlocal behavior in the microscopic world. These ideas are generalized
further to noncommutative space to remove singularity in particle physics
and to understand the nonlocality of gravity and quantum cosmology. \cite{Douglas} 
The figure 1 shows the family of noncommutative space. The
Heisenberg's space is defined by
\begin{eqnarray}
\left[ x_{i},p_{j}\right] &=&i\hbar \delta _{ij}\text{ \ } \\
\left[ x_{i},x_{j}\right] &=&\left[ p_{i},p_{j}\right] =0
\end{eqnarray}%
where $i,j=1,2,3$. This is also called as a canonical commutative relation
or Heisenberg's algebra. This noncommutative algebra can be generalized
to \cite{Seiberg}
\begin{eqnarray}\label{xp2}
\left[ \widehat{x}_{i},\widehat{p}_{j}\right] &=&i\hbar \Delta _{ij} \\
\left[ \widehat{x}_{i},\widehat{x}_{j}\right] &=&i\theta _{ij} \\
\left[ \widehat{p}_{i},\widehat{p}_{j}\right] &=&i\widetilde{\theta }_{ij}
\end{eqnarray}%
where $\theta _{ij}$ and $\widetilde{\theta }_{ij}$ are antisymmetric real
constant $3\times 3$ matrices. These noncommutative algebra are equivalent
to the Weyl-Moyal correspondence in field theories on noncommutative spaces
\cite{Douglas} and can be realized by the Seiberg-Witten map, \cite{Seiberg}
\begin{eqnarray}\label{xp3}
\widehat{x}_{\mu } &=&a_{\mu \nu }x_{\nu }+b_{\mu \nu }p_{\nu } \\
\widehat{p}_{\mu } &=&c_{\mu \nu }x_{\nu }+d_{\mu \nu }p_{\nu }  
\end{eqnarray}
between the canonical commutative relation and noncommutative phase space.
For two-dimensional system, the noncommutative matrices can be obtained by the Seiberg-Witten map
\begin{equation}
\Delta _{ij} =\left(
\begin{array}{cc}
\alpha^2+\frac{\theta \widetilde{\theta }}{2\alpha^2 \hbar^2 } & \frac{\theta \widetilde{\theta }}{4\alpha^2 \hbar^2 } \\
\frac{\theta \widetilde{\theta }}{4\alpha^2 \hbar^2 } & \alpha^2+\frac{\theta \widetilde{\theta }}{2\alpha^2 \hbar^2 } \\
\end{array}%
\right)
\end{equation}
with
\begin{equation}
\theta _{ij} =\frac{1}{\hbar }\left(
\begin{array}{cc}
0 & \theta \\
-\theta & 0 \\
\end{array}%
\right);\qquad  
\widetilde{\theta } _{ij} =\frac{1}{\hbar }\left(
\begin{array}{cc}
0 & \widetilde{\theta } \\
-\widetilde{\theta } & 0 \\
\end{array}%
\right)
\end{equation}
If we may set the constraint of $\alpha$ and $(\theta,\widetilde{\theta })$,
$\alpha^2+\frac{\theta \widetilde{\theta }}{2\alpha^2 \hbar^2 }=1$ such that
$\Delta _{ij} =\left(
\begin{array}{cc}
1 & \frac{1-\alpha^2}{2} \\
\frac{1-\alpha^2}{2} & 1 \\
\end{array}%
\right)$.
When $\alpha=1$, $\Delta _{ij}=\delta _{ij}$ and $\theta\widetilde{\theta }=0$.

\begin{figure}[tbp]\label{fig1}
\centering
\includegraphics[scale=0.4]{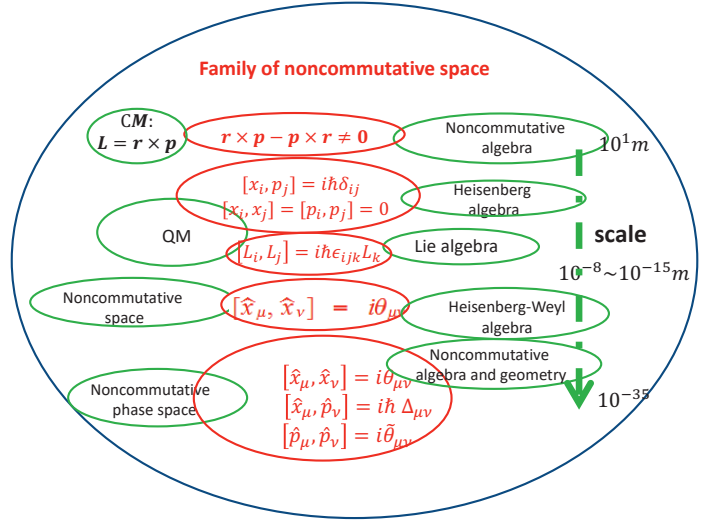}
\caption{ (Color online): The family of noncommutative space.}
\end{figure}

Interestingly, Connes and Rieffel construct a mathematical formulation on
noncommutative geometry and algebra, which provide a mathematical language
of noncommutative space physics.\cite{Connes}
However, so far there has not been direct experimental evidences to
demonstrate the existence of noncommutative geometry or space even
though a lot of phenomena in particle physics and gravity can be described
by the noncommutative space or phase-space language. The difficulty to
directly find out experimental evidence of the existence of noncommutative
space is because the noncommutative space phenomena are predicted occurring
in the Planck scale.

In this paper, we apply the Seiberg-Witten map to study the Aharnov-Bohm
effect in noncommutative phase space. The noncommutative effect induces an
effective magnetic flux that produces the persistent current in a mesoscopic
ring. We propose a scheme to probe the effective magnetic flux coming from
the noncommutative phase space even the effect is very weak. Since the
persistent current and magnetic flux in mesoscopic rings can be implemented
by nanotechnology, it can be expected to detect the noncommutative phase
space effect by this scheme.

\section{Quantum Ring in noncommutative phase space}

The Hamiltonian of the two-dimensional (2D) free electron in noncommutative
phase space can be written as
\begin{equation}
H_{nc}=\frac{1}{2m}\left( \widehat{p}_{x}^{2}+\widehat{p}_{y}^{2}\right)
\end{equation}%
which can be mapped to the Heisenberg's (canonical) commutative space
\begin{equation}
H_{nc}=\frac{1}{2m^{\ast }}\left[ \left( p_{x}+eA_{x}\right) ^{2}+\left(
p_{y}+eA_{y}\right) ^{2}\right]
\end{equation}%
where $m^{\ast }=m/\alpha $\ is the effective mass and the effective vector
potential is
\begin{equation}
A_{x}=\frac{\widetilde{\theta }}{2e\alpha ^{2}\hbar }y;\text{ \ \ \ }A_{y}=-%
\frac{\widetilde{\theta }}{2e\alpha ^{2}\hbar }x
\end{equation}%
and the effective magnetic field

\begin{equation}
B_{z}=\frac{\widetilde{\theta }}{e\alpha ^{2}\hbar }
\end{equation}%
It can be seen that the free electron in the 2D noncommutative phase space
is equivalent to electron in an effective magnetic field induced by the
noncommutative phase space effect, which provides a way to detect the
existence of noncommutative phase space by probing the effect of the
effective magnetic field or vector potential coming from the noncommutative
phase space.

\begin{figure}[tbp]\label{fig2}
\centering
\includegraphics[scale=0.4]{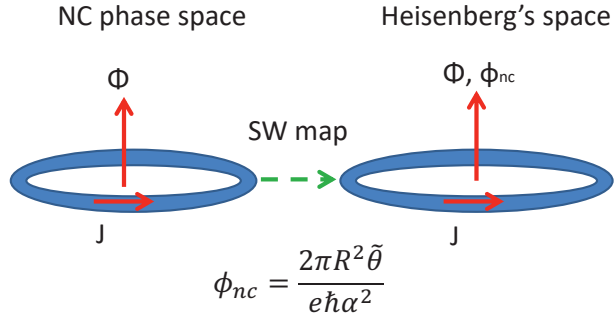}
\caption{ (Color online): The nano scale ring in the noncommutative phase space on the left side, which is mapped (Seiberg-Witten) to 
Heisenberg's (canonical) space on the right side.}
\end{figure}

We consider a one-dimensional ring in an external magnetic field along the
ring axis. The magnetic field is constant inside the ring such that the
electron states depend only on the total magnetic flux in the ring. In the
polar coordinate system, $x=R\cos \varphi ,y=R\sin \varphi $, by using the
Seiberg-Witten map, the Hamiltonian in noncommutative phase space is written
as 

\begin{equation}
H_{nc}=-\frac{\hbar ^{2}}{2m^{\ast }R^{2}}\left[ \frac{\partial }{\partial
\varphi }+i\left( \frac{\phi }{\phi _{0}}-\frac{\phi _{nc}}{\phi _{0}}%
\right) \right] ^{2}-\frac{3\hbar ^{2}}{8m^{\ast }R^{2}}\frac{\phi _{nc}^{2}%
}{\phi _{0}^{2}}
\end{equation}%
where $\phi _{nc}=\frac{2\pi R^{2}\widetilde{\theta }}{e\hbar \alpha ^{2}}$
is the effective magnetic flux coming from the noncommutative phase space
and $\phi _{0}=\frac{h}{e}$ is the flux quanta ($e<0$). $\phi $ is the external
magnetic flux in the ring. (see Fig. 2) For convenience, we introduce the
dimensionless magnetic flux, $f_{nc}\equiv \frac{\phi _{nc}}{\phi _{0}}$ and
$f=\frac{\phi }{\phi _{0}}$, the Hamiltonian of quantum ring can be
rewritten as 
\begin{equation}
H_{nc}=-\varepsilon _{0}\left[ \frac{\partial }{\partial \varphi }+i\left(
f-f_{nc}\right) \right] ^{2}-\frac{3\varepsilon _{0}}{4}f_{nc}^{2}
\end{equation}%
where $\varepsilon _{0}\equiv \frac{\hbar ^{2}}{2m^{\ast }R^{2}}$. The
eigenenergy can be obtained by \cite{Liang}
\begin{equation}
E_{n}=\varepsilon _{0}\left( n+f-f_{nc}\right) ^{2}-\frac{3\varepsilon _{0}}{%
4}f_{nc}^{2}  \label{Eign}
\end{equation}%
where $n=0,\pm 1,\pm 2,...$. It can be seen that the effective magnetic flux
induced by the noncommutative space modifies the eigenenergy levels. Since $%
n=0,\pm 1,\pm 2,..$, and notice that the eigenenergies in Eq. (\ref{Eign})
are invariant under $f-f_{nc}\rightarrow f-f_{nc}+1$, we can consider only
the domain of $f-f_{nc}$ within $\left[ -\frac{1}{2},\frac{1}{2}\right] $
(the first Brillouin flux zone).\cite{Daniel,Huang} Suppose there are $N$
electrons in the ring, and they occupy the energy levels in zero
temperature, notice that for the odd-electron ring, $N=2k+1$, the
ground-state energy is $E_{g}=\sum_{n=0,\pm 1,\pm 2,...}^{\pm k}E_{n}$, and
for the even-electron ring, $N=2k$, $E_{g}=\left(
\sum_{n=0,1,2,...}^{k-1}+\sum_{n=-1,-2,...}^{-k}\right) E_{n}$, the
ground-state energy of the ring is obtained \cite{Liang}

\begin{equation}\label{Eg}
E_{g}=\varepsilon _{0}\left\{
\begin{array}{cc}
\frac{N^{3}-N}{12}+N\left[ (f-f_{nc})^{2}-\frac{3}{4}f_{nc}^{2}\right] &
for\ N=2k+1 \\
\frac{N^{3}+2N}{12}-N(f-f_{nc})+N\left[ (f-f_{nc})^{2}-\frac{3}{4}f_{nc}^{2}%
\right] & for\ N=2k%
\end{array}%
\right.  
\end{equation}%
Moreover, the ground-state energy is symmetric for $f-f_{nc}=0$, we can
restrict our attention to a half of the first Brillouin flux zone, $\left[ 0,%
\frac{1}{2}\right] $. The persistent current in the ground state is defined
by $J=-\frac{\partial E_{g}}{\partial \phi }$ and it can be obtained \cite{Liang}

\begin{equation}\label{Jphi}
J=J_{0}\left\{
\begin{array}{cc}
-2N\frac{\phi }{\phi _{0}}\left( 1-\frac{\phi _{nc}}{\phi }\right) & for\
N=2k+1 \\
N-2N\frac{\phi }{\phi _{0}}\left( 1-\frac{\phi _{nc}}{\phi }\right) & for\
N=2k%
\end{array}%
\right.
\end{equation}
where $J_{0}=\frac{e}{h}\varepsilon _{0}$. The persistent current depends on
both of the external magnetic flux and the effective magnetic flux induced
by the noncommutative phase space. This relationship between the persistent
current and the magnetic flux in Eq. (\ref{Jphi}) provides a way to reveal
the noncommutative phase space effect.

\section{Signatures of noncommutative-phase-space effect}

In order to detect the noncommutative-phase-space effect experimentally, we
introduce two variables defined by \cite{Liang}
\begin{eqnarray}
\lambda&\equiv& \frac{\partial }{\partial \phi }\left(\frac{J}{\phi}\right)
\\
\sigma&\equiv& \frac{\partial }{\partial \phi }\left(\frac{J-NJ_{0}}{\phi }%
\right)
\end{eqnarray}
as two signatures to detect the noncommutative-phase-space effect
experimentally. Thus, we get

\begin{eqnarray}
\lambda=\left\{
\begin{array}{cc}
-2NJ_{0}\frac{f_{nc}}{\phi ^{2}} & for\ N=2k+1 \\
-NJ_{0}\left(1+2f_{nc}\right)\frac{1}{\phi^{2}} & for\ N=2k%
\end{array}
\right.
\end{eqnarray}
and
\begin{eqnarray}
\sigma=\left\{
\begin{array}{cc}
NJ_{0}\left(1-2f_{nc}\right)\frac{1}{\phi^{2}} & for\ N=2k+1 \\
-2NJ_{0}\frac{f_{nc}}{\phi ^{2}} & for\ N=2k%
\end{array}
\right.
\end{eqnarray}
It can be seen that when there exists the noncommutative phase space, both $%
\lambda $ and $\sigma $ are proportional to $\frac{\pm 1}{\phi ^{2}}$, which
diverge in the small external magnetic fluxes.

For given parameter $\widetilde{\theta }\leq 1.76\times
10^{-61}[kg]^{2}[m]^{2}[s]^{-2}$ \cite{Bastos} and a mesoscopic ring with
radius $R=1\mu m$, $f_{nc}\leq \frac{2\pi R^{2}\widetilde{\theta }}{e\hbar
\alpha ^{2}h/e}=\frac{R^{2}\widetilde{\theta }}{\hbar ^{2}\alpha ^{2}}%
=\allowbreak 1.5828\times 10^{-5}$. Suppose that the effective electron
number in the ring is about $10^{4}\sim 10^{5}$, we show some theoretical
predictions in Figures 3 and 4 based on this idea and parameters. The figure
3 (a) shows the theoretical prediction of the $\lambda \sim \phi $ behavior
in nano-ring with radius $R=1\mu m$ with odd-electron number. It can be seen
that the $\lambda $ is divergent in the small $\phi $ range. Similarly in
Fig. 3 (b), we shows the behavior of $\sigma \sim \phi $. It can be seen
that both $\lambda $ and $\sigma $ for odd electron number rings have a
similar behavior, but the sign is different.

For the even-electron number rings, the behaviors of $\lambda $ and $\sigma $
are similar and have only a two-order difference, which are shown in Fig. 4.

\begin{figure}[tbp]\label{fig3}
\centering
\includegraphics[scale=0.5]{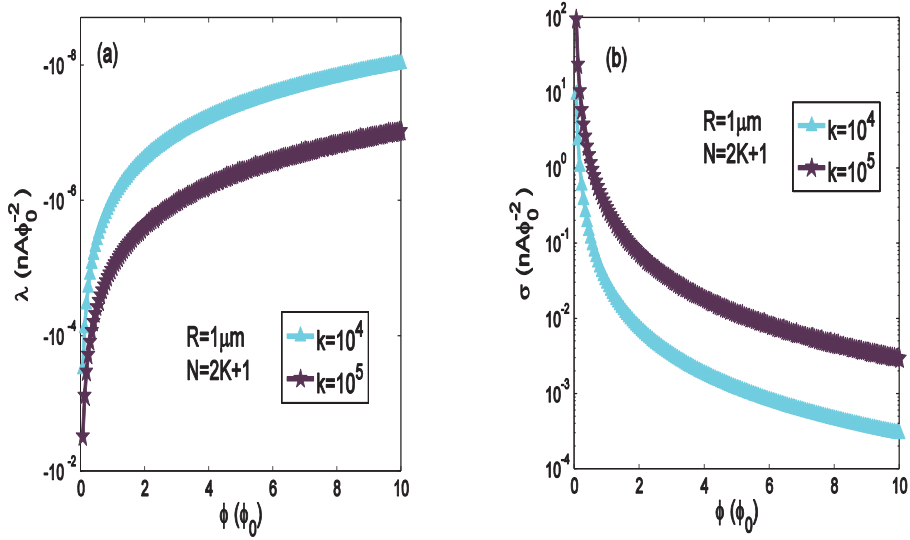}
\caption{ (Color online): The signatures of noncommutative phase space for
even electron-number rings versus the external magnetic flux. (a) $\log%
\protect\lambda\sim\protect\phi$ and $\log\protect\sigma\sim\protect\phi$
with different electron numbers. (b) $\log\protect\lambda\sim\log\protect%
\phi $ and $\log\protect\sigma\sim\log\protect\phi$ a standard of $1/\protect%
\phi$ behavior. The unit of the magnetic flux is the magnetic flux quantum $%
\protect\phi_{0}$.}
\end{figure}

\begin{figure}[tbp]\label{fig4}
\centering
\includegraphics[scale=0.5]{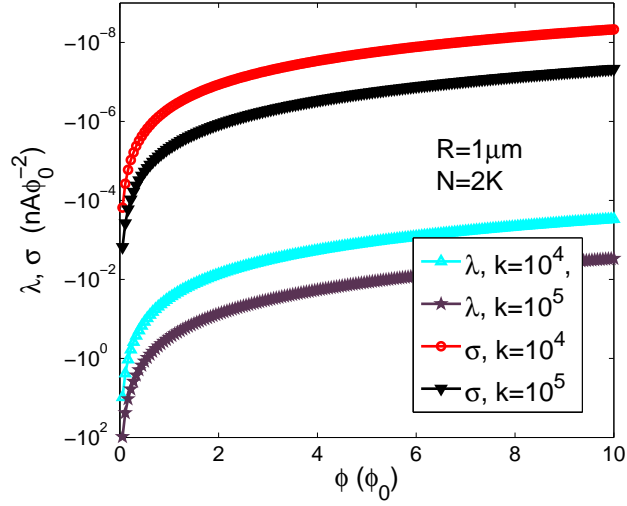}
\caption{ (Color online): 
The signatures of noncommutative phase space for even-electron number rings versus the external magnetic flux.
$\log\protect\lambda\sim\protect\phi$ and $\log\protect\sigma\sim\protect\phi$ with different electron numbers.  The unit of
the magnetic flux is the magnetic flux quantum $\protect\phi_{0}$.}
\end{figure}

\section{Criterion and Scheme for detecting noncommutative-phase-space effect}

We can give a criterion to directly detect the existence of the
noncommutative phase space.\newline
\textbf{Criterion}: if one of the following two cases occurs it infers the
existence of the noncommutative phase space:\newline
(1) $\lambda \sim \phi $ is $-1/\phi ^{2}$ divergent but $\sigma \sim \phi $
is $1/\phi ^{2}$ divergent (see Fig. 3);\newline
(2) both $\lambda \sim \phi $ and $\sigma \sim \phi $ are $-1/\phi ^{2}$
divergent and $\lambda <\sigma $ (see Fig. 3).

If (1) occurs it infers that there is odd number of electrons in the ring
and (2) means the ring having even number of electrons.

Based on this criterion, we propose an experimental scheme to demonstrate
explicitly the existence of the noncommutative phase space. \newline
\textbf{Experimental Scheme}: The basic steps include

(1) setting up a mesoscopic ring system with an external magnetic field;

(2) measuring the persistent current $J$ versus the external magnetic flux $%
\phi $;

(3) calculating $\lambda $ and $\sigma $ by using the numerical
interpolation and derivative techniques for estimating the electron number $N
$;

(4) plotting $\lambda $ and $\sigma $ versus $\phi $, if we can
obtain a qualitative the behaviors of $\lambda $ versus $\phi $ or $\sigma $
versus $\phi $ in Figs 3 and 4, they demonstrate the existence of the
noncommutative phase space.

It should be emphasized that the resolution of the numerical estimations in
Figs 3 and 4 are based on the parameters of the Planck scale and nano size
of the ring. The comparison between the experimental data and the
theoretical prediction in Figs 3 and 4 are meaningful and valid for all
qualitative level because  actually we do not know the exact values of these
parameters. In other words, the divergent behavior $\pm 1/\phi ^{2}$ from the
experimental data provides a way to predict the values of the noncommutative
parameter $\widetilde{\theta }$. In practice, we can plot the $\log \lambda
\sim\log \phi $ such that we can verify the divergent behavior $\pm
1/\phi ^{2}$ as a linear relation.

In fact, the persistent current in mesoscopic ring have been studied both
theoretically and experimentally in the past two decades. \cite{
Buttiker, Levy, Mailly} Buttiker first predicted that the persistent current
occurs in mesoscopic ring and oscillates with an AB flux.\cite{Buttiker} The
amplitude of the persistent current reaches $\left( 10^{-2}\sim 2\right)
ev_{F}/2\pi R$ where $v_{F}$ is the electronic velocity at Fermi level in
the $Cu$ multi-ring system, an isolate $Au$ ring and $GaAs/Al_{x}Ga_{1-x}As$
in the diffusive region at low temperature.\cite{Levy,Chandrasekhar, Mailly}%
, which agrees with the theoretical prediction. This  nanotechnology
provides an efficient way to probe the noncommutative phase space. We
strongly suggest to rerun the persistent-current experiment to turn out the
persistent versus the external magnetic flux data, by which we can
investigate the relationship between $\left( \lambda ,\sigma \right) $ and $%
\phi $ for detecting the existence of the noncommutative phase space.

On the other hand, Carroll et. al. studied the noncommutative field theory
and Lorentz violation. They gave an upper bound of the noncommutative
parameter, $\widetilde{\theta }\leq \left( 10TeV\right) ^{-2}$.\cite{Carroll}
Falomir et. al. also proposed a scheme to explore the spatial
noncommutativity of the scattering differential cross section by the AB
effect.\cite{Falomir} It relies on the particle physics experiment involving
energies between $200$ and $300$ GeV for $\widetilde{\theta }\leq \left(
10TeV\right) ^{-2}$ and estimating the typical order of the cross section
for neutrino events $10^{-3}$. \cite{Falomir} Obviously, that
experimental schemes are much more difficult to be implemented than this
scheme because this experimental scheme involves $eV$ energy scale and
nanoscale physics.

In fact, there has not been any direct experimental evidences to demonstrate
the existence of noncommutative space and phase space, it should be worth
studying and exploring from different aspects and different energy scales
even though the concept of noncommutativity originates from the Planck scale
physics. Actually, many phenomena in condensed matter physics show the
characteristics of noncommutative space in nonrelativistic quantum
mechanics, such as an analogy between the Landau's levels of two-dimensional
electron gas in the presence of magnetic field and the free electron in
noncommutative phase space,\cite{Gamboa,Ho}, as well as quantum Hall effect.%
\cite{Basu}, it should be expected some possibilities to catch the
noncommutativity especially in the nanoscale condensed matter physics.

\section{Conclusions}

In summary, we propose an experimental scheme to probe the existence of
noncommutative phase space by the nano-scale quantum ring based on the
Seiberg-Witten map. We introduce two variables $\lambda $ and $\sigma $ as
two signatures to detect the effect of the noncommutative phase space based
on the relationship between the persistent current and the external magnetic
flux in the ring. The divergent behaviors of  $\left( \lambda ,\sigma
\right) $ versus $\phi $ provide a value-independent scheme for the
experimental measurement to catch the effect of the noncommutative phase
space. It can be expected to give the answer whether exists the
noncommutative phase space.

\begin{figure}[tbp]\label{fig6}
\centering
\includegraphics[scale=0.4]{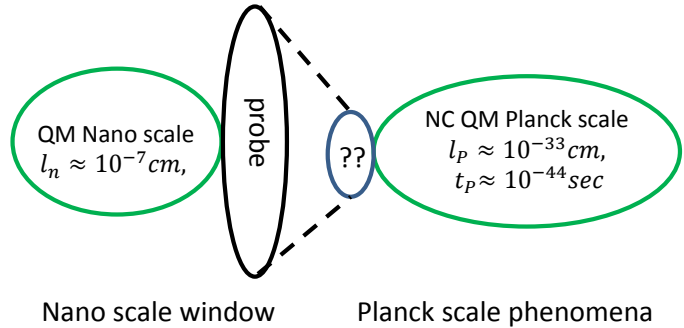}
\caption{ (Color online): The nano scale window for probing the Planck scale phenomena.}
\end{figure}

The noncommutative Heisenberg's algebra plays a crucial role in quantum
mechanics. It provides an efficient way to understand many novel phenomena
beyond classical physics, such as the wave-particle duality of elementary
particles, uncertainty of a pair conjugated variables, quantization of
physical variables, and nonlocal quantum entanglement. These novel
properties in microscopic world imply that the noncommutative Heisenberg's
algebra could be extended to the noncommutative spacetime or phase space to
avoid the singularity in particle physics, gravity and early universe in the
Planck scale world. It should be a meaningful direction to explore the
noncommutativity of space or phase space in the mesoscopic scale world
because it is too difficult to directly implement physical experiments in
Planck scale. This scheme gives a nano scale window to explore
noncommutativity in the Planck's scale world.


\begin{acknowledgments}
The author thanks the financial support of the Natural Science Fundation of
Guangdong Province (No. 2016A030313313).
\end{acknowledgments}


\bibliographystyle{plain}
\bibliography{apssamp}

\begin{references}

\bibitem{Douglas} Michael R. Douglas, Nikita A. Nekrasov, Rev. Mod. Phys. 73, (2001)977;
Richard J. Szabo, Physics Reports 378 (2003) 207; H. Snyder, Phys. Rev. 71, (1947)38.

\bibitem{Seiberg} Seiberg, N., and E. Witten,  Nucl. Phys. B 426, (1994)19; Seiberg, N., and E. Witten,
J. High Energy Phys. 09, (1999)032.

\bibitem{Connes} A. Connes, Noncommutative Geometry (Academic, San Diego, 1994).

\bibitem{Liang} Shi-Dong Liang, Haoqi Li and Guang-Yao Huang, Phys. Rev. A 90, 010102 (2014);
Shi-Dong Liang, Poster presentation on Conference of 90 Years of Quantum Mechanics, NTU, Singapore, Jan. 2017.

\bibitem{Bastos} C. Bastos , O. Bertolam, Phys. Letters A 372 (2008) 5556;
O. Bertolami, J.G. Rosa, C. Aragao, P. Castorina, D. Zappal¨¤, Phys. Rev. D 72 (2005) 025010.

\bibitem{Daniel}Daniel Loss and Paul Goldbart, Phys. Rev. B 43, (1991)13762.

\bibitem{Huang} Guang-Yao Huang, Shi-Dong Liang, Phys. Lett. A 375, (2011)738.

\bibitem{Buttiker}M. Buttiker, Y. Imry, and R. Landauer, Phys. Lett. 96A (1983) 365.

\bibitem{Levy} L. P. Levy, G. Dolan, J. Dunsmuir, and H. Bouchiat, Phys. Rev. Lett.
64, (1990)2074.

\bibitem{Chandrasekhar} V. Chandrasekhar, R. A. Webb, M. J. Brady, M. B. Ketchen, W. J.
Gallagher, and A. Kleinsasser, Phys. Rev. Lett. 67, (1991)3578.

\bibitem{Mailly} D. Mailly, C. Chapelier, and A. Benoit, Phys. Rev. Lett. 70, (1993)2020.

\bibitem{Carroll} S. Carroll, J. Harvey, V. A. Kostelecky, C. D. Lane and T. Okamoto, Phys. Rev. Lett. 87, (2001)141601.

\bibitem{Falomir} H. Falomir, J. Gamboa, M. Loewe, F. Mendez and J. C. Rojas, Phys. rev. D 66, (2002)045018.


\bibitem{Gamboa} J. Gamboa, M. Loewe, J.C. Rojas, Phys. Rev. D 64 (2001) 067901;
M. Chaichian, M.M. Sheikh-Jabbari, A. Tureanua, Phys. Rev. Lett. 86 (2001).

\bibitem{Ho} P.-M. Ho, H.-C. Kao, Phys. Rev. Lett. 88 (2002) 151602.

\bibitem{Basu}B. Basu, Debashree Chowdhury, SubirGhosh, Phys. Lett. A 377 (2013) 1661.


\end{references}


\end{document}